\documentclass[prl,aps,preprint,amsmath,amssymb]{revtex4}
\usepackage{graphicx}
\pdfoutput=1
\usepackage{amsmath}
\usepackage{color}
\usepackage{float}

\begin{document}

\title{High-temperature ferroelectricity in SrTiO$_{3}$ crystals}
\author{Ashima Arora}
\author{Goutam Sheet}
\email{goutam@iisermohali.ac.in}
\affiliation{Department of Physical Sciences,  
Indian Institute of Science Education and Research (IISER), Mohali,
Punjab, India, PIN: 140306}

\begin{abstract}
SrTiO$_{3}$ is known to be an incipient ferroelectric. It is thought that ferroelectric stability in SrTiO$_{3}$ is suppressed by a delicate competition with quantum fluctuation and antiferrodistortion. The ferroelectric phase can, however, be stabilized by doping, isotope manipulation and strain engineering etc. Till date ferroelectricity in SrTiO$_{3}$ thin films was observed to exist up to room temperature -- that was when the films were grown on specially engineered substrates using complex growth techniques. It was possible to write and erase ferroelectric domains on the specially engineered films at room temperature. Here, we show remarkably similar ferroelectric behavior in bulk (110) single crystals of  SrTiO$_{3}$ with no special engineering  well above room temperature using piezoresponse force microscopy. Hysteretic switching of local electric polarization was observed and electrically active domains could be written and erased using lithographic techniques at remarkably high temperatures up to $420K$. 

\end{abstract}

\maketitle

Strontium titanate (SrTiO$_{3}$) has found wide range of application as a substrate for the epitaxial growth of functional oxide thin films because of its high dielectric constant.\cite{Dawber} The list of oxides that have been successfully grown on SrTiO$_{3}$ with high degree of epitaxy includes high temperature ceramic superconductors, colossal magnetoresitive manganites, itinerant ferromagnets etc.\cite{Ohtomo, Ahn,Fong} A large number of devices have also been fabricated on SrTiO$_{3}$.\cite{Burnside, Tumarkin} More recently, it has been shown that a two-dimensional electron gas is formed under certain conditions at the interface of insulating SrTiO$_{3}$ and insulating LaAlO$_{3}$ (LAO) where novel interplay between superconductivity and magnetism has been reported.\cite{Ohtomo, Manan} Therefore, the bulk and surface properties of SrTiO$_{3}$ have attracted considerable attention in contemporary condensed matter physics. 

SrTiO$_{3}$ crystallizes in the cubic perovskite structure in its paraelectric phase and is known to lead to a cubic to antiferrodistortive phase transition at $105K$.\cite{Rimai, Bell, Zhong} From first principle calculations, it was inferred that SrTiO$_{3}$ might possibly have a low temperature $(40K)$ ferroelectric phase under hydrostatic pressure or strain.\cite{Antons} As the ferroelectric phase was not experimentally detected it was believed that quantum fluctuations forbade a stable ferroelectric phase at low temperatures.\cite{Zhong,Muller} The idea of quantum fluctuations was further supported by the observation of a ferroelectric phase when the oxygen of SrTiO$_{3}$ was replaced with the $^{18}O$ isotope.\cite{Itoh} However, more recently, it has been shown that ferroelectricity in epitaxial thin films of SrTiO$_{3}$ could be induced by strain engineering.\cite{Haenl,Bilani-Zeneli,Kim,Maeng} Nevertheless, the consensus has been that the ferroelectric transition cannot be observed in pure bulk crystals of SrTiO$_{3}$. However, more recently, by superior thin film deposition methods a room temperature ferroelectric phase of strain-free SrTiO$_{3}$ film could be achieved.\cite{Jang}  The observation of ferroelectricity in strain free SrTiO$_{3}$ naturally leads to the idea that room temperature ferroelectricity might, in principle, be possible in bulk single crystal of SrTiO$_{3}$ as well. In fact, from computer simulation \cite{Ravikumar}, low-energy electron diffraction\cite{Bickel} and surface x-ray diffraction\cite{Herger} it was argued that surface ferroelectricity in a few monolayers of bulk SrTiO$_{3}$  should be possible. In this Letter,we show that the surface of SrTiO$_{3}$ shows hysteretic switching of polarization with DC bias. It is also possible to write and erase ferroelectric domains on single crystals of SrTiO$_{3}$ at remarkably high temperatures up to $420K$ indicating the existence of ferroelectricity in bulk SrTiO$_{3}$ at such high temperatures.

\begin{figure*}

\includegraphics[width=1\textwidth]{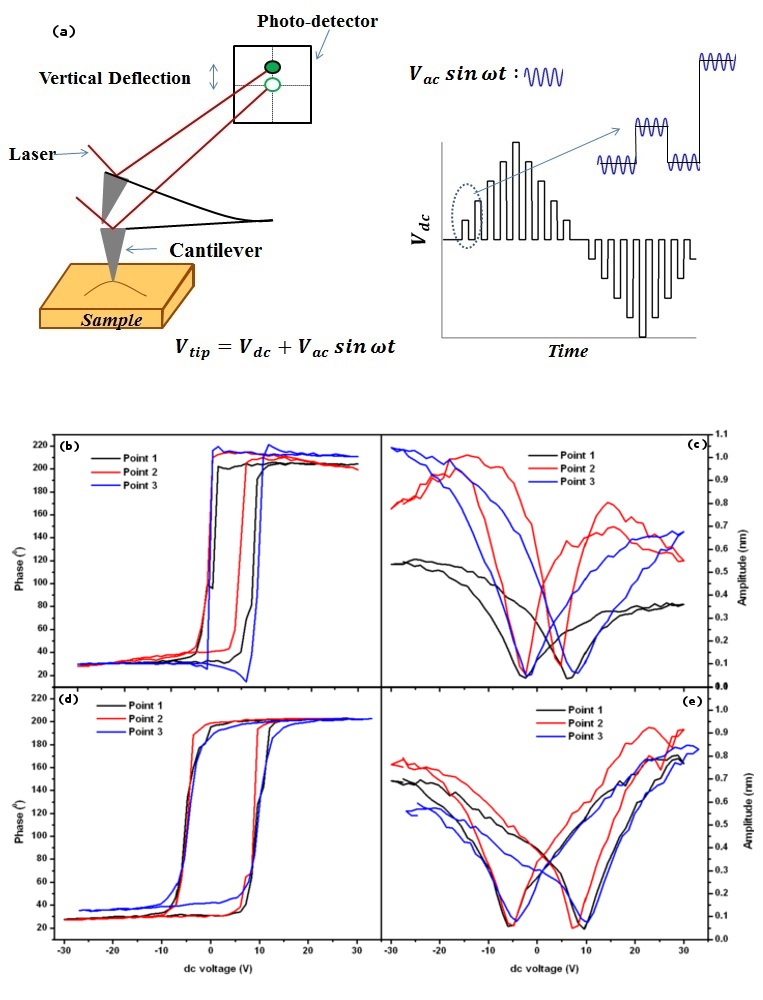}

\caption{(a) Schematic description of the PFM technique. The switching waveform in the DART-PFM switching spectroscopic mode is also shown.(b) PFM phase hysteresis and (c) butterfly loops at $300K$ at three different points in the ``off" state. (d) PFM phase hysteresis and (e) butterfly loops at $420K$ at three different points in the ``off" state as in (d).}
\end{figure*}

In order to probe ferroelectricity in bulk single crystals of SrTiO$_{3}$, we employed piezoresponse force microscopy (PFM) and switching spectroscopy. In this measurement a cantilever with a conducting tip is mounted in an atomic force microscope (AFM) and is electrically connected to a DC high-voltage (up to $220V$) amplifier and a lock-in amplifier. The lock-in amplifier sends a sinusoidal voltage $(V_{ac}sinwt)$ to the cantilever, which is brought in contact with the sample during PFM measurements. The amplitude response from the cantilever in the contact mode is probed as function of $w$ in order to find the in-contact resonance frequency $(w_{r})$ of the cantilever. The measurements are performed at $w_{r}$ in order to obtain maximum sensitivity. It is believed that a $180^{0}$ switching of the phase of the response signal from the cantilever is a signature of ferroelectricity. The switching behaviour should also be hysteretic when a DC voltage $(V_{dc})$ is swept on the cantilever. This is attributed to the switching of the ferroelectric domains in the ferroelectrics. On the other hand, hysteretic amplitude vs. voltage curve, also known as a ``butterfly loop", is the hallmark of piezoelectricity.

\begin{figure*}
\begin{center}
\includegraphics[width=1\textwidth]{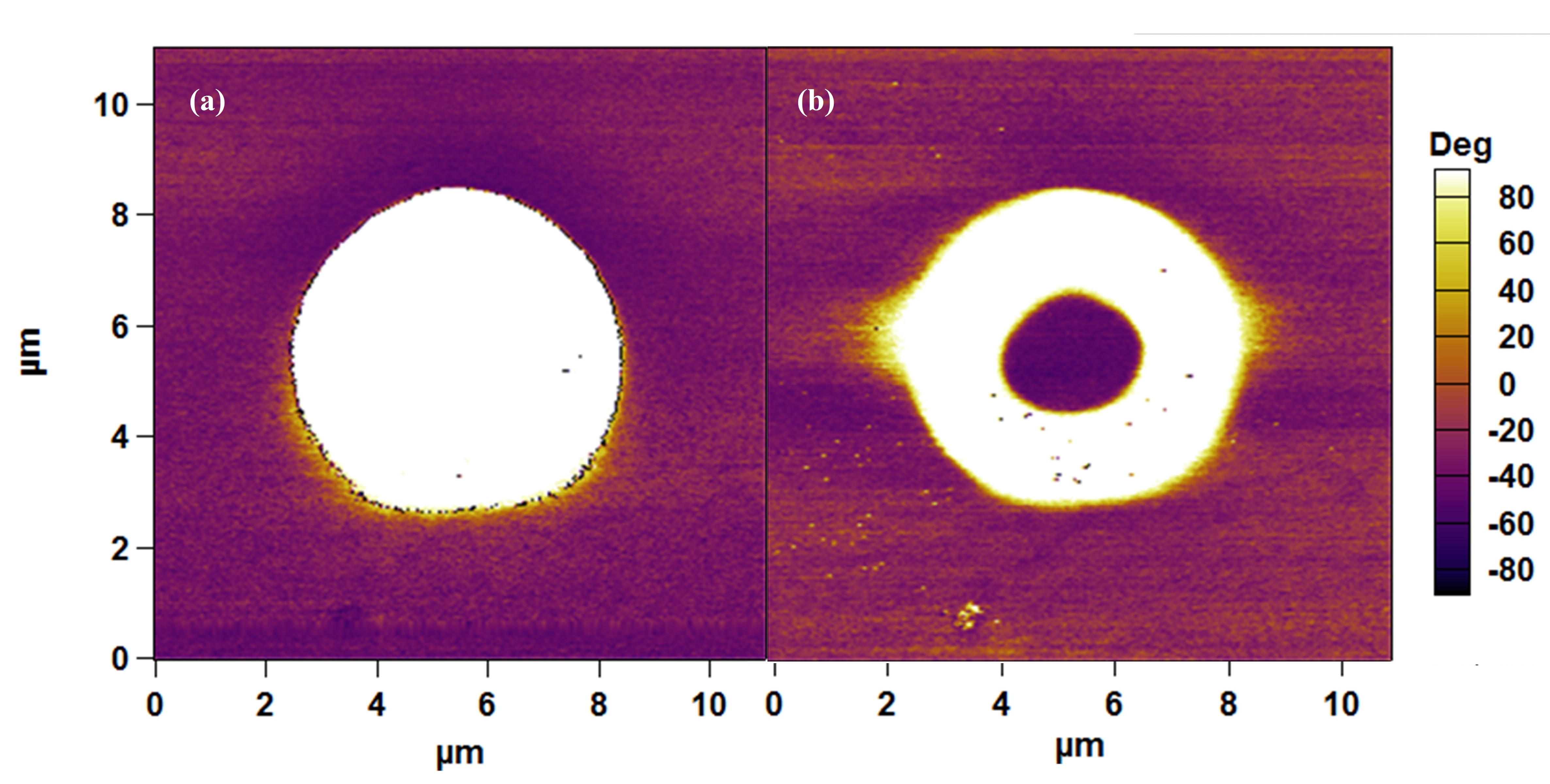}
\end{center}

\caption{ (a) Phase image of a circular domain written at room temperature by applying $+30V$ and (b) phase image of the domain after applying $-30V$ at the center of the domain written in (a).}
\end{figure*}

A (110) single crystal of SrTiO$_{3}$ (from MTI Corporation) was cut into a $3mm \times 3mm$ piece and was loaded on a metallic sample holder. The sample holder was mounted on the sample stage of an AFM where the metallic sample holder was internally connected to the ground of the high voltage amplifier. In order to measure temperature dependence of the data, we had mounted a heating kit in the AFM. The sample was in direct thermal connection with the hot surface of the heating kit. The heating kit was capable of raising the sample temperature up to $570K$. The cantilevers that were used for these measurements were made of silicon and coated with $5nm$ of titanium and $20nm$ of iridium. The spring constant of the cantilever was $2N/m$ with its natural resonance frequency in air around $70kHz$. The in-contact resonance frequency of the cantilever on SrTiO$_{3}$ was found to be in the range of $290-310kHz$. 

\begin{figure*}
\begin{center}
\includegraphics[width=1\textwidth]{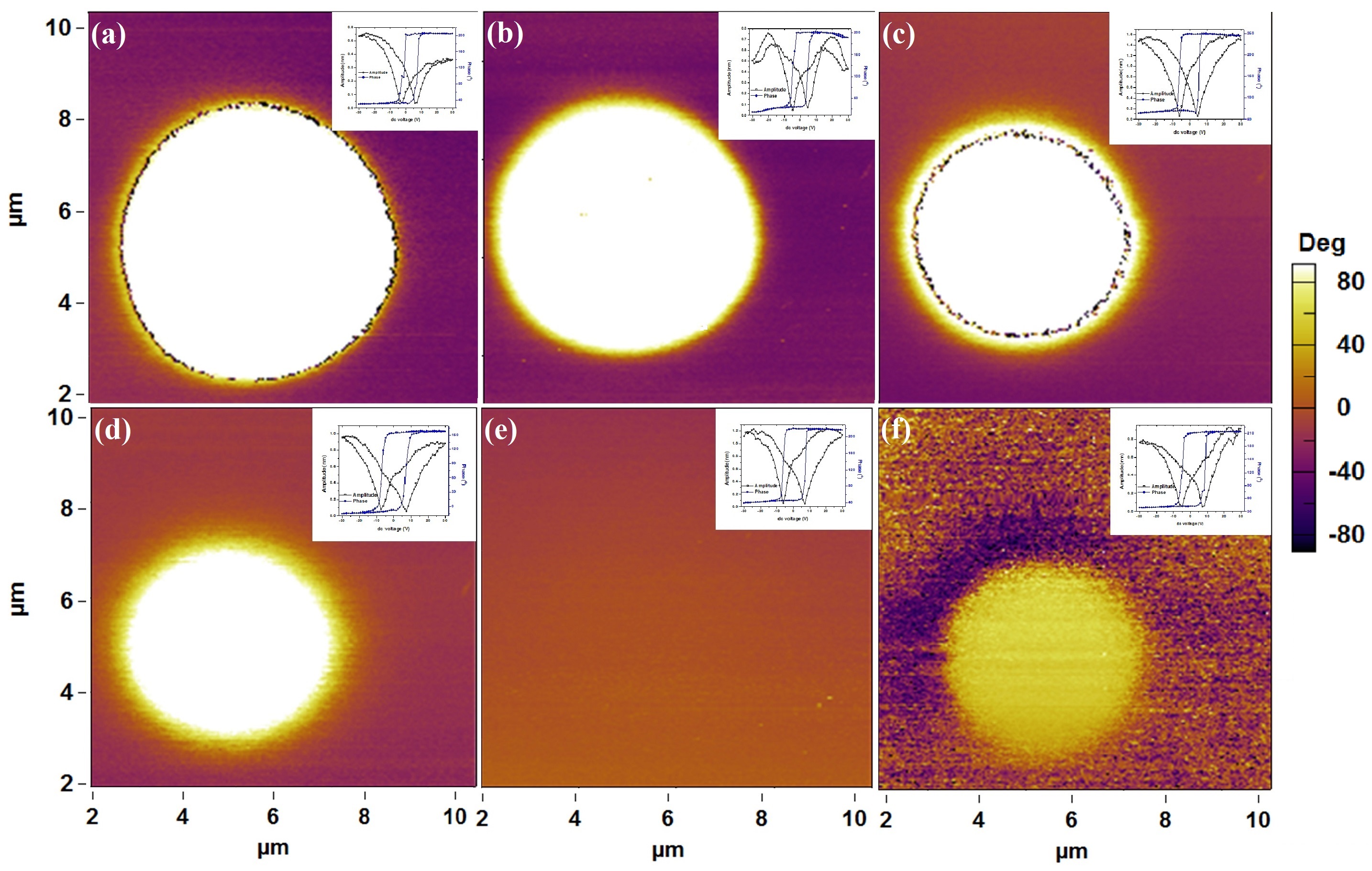}

\caption{(a) A circular domain written on bulk SrTiO$_{3}$ with $+30V$ on a $8 \mu m \times 8 \mu m$ area at $300K$. The PFM Phase images of the domain at different temperatures: (b) $330K$ (c) $360K$ (d) $390K$ (e) $420K$; (f) Phase image (at $300 K$ of the domain written with $+30V$ at $440K$ and then immediately cooling the sample down to $300K$.}
\end{center}
\end{figure*}
It should be noted that hysteretic phase switching and butterfly loops may also originate from reasons other than ferroelectricity and piezoelectricity in PFM.\cite{Sekhon, Proksch} One of the possible reasons could be the local electrostatic effects. In order to mitigate this effect, all the spectroscopic data reported here were measured by employing a switching spectroscopy PFM (SSPFM) protocol pioneered by Jesse \textit{et.al.}, where a sequence of DC voltages in triangular saw tooth form was applied between the conducting cantilever and the SrTiO$_{3}$ as shown in Figure 1(a).\cite{JesseA, JesseB} The measurement was carried out in the ``off" state of the pulses. As shown in the supplementary material\cite{EPAPS}, there is a significant difference between the ``on" state and the ``off" state data. This suggests that the local electrostatic effects have been minimised during the spectroscopic measurements. The electro-chemical response in the presence of a high DC bias may also mimic the ferroelectric like hysteresis effects. Usually one clear signature of an electrochemical reaction is the formation of topographic structures on the surface.\cite{Sekhon, Proksch} However, in the present case, no topographic structure growth was noticed (see supplementary material).\cite{EPAPS} 

In Figure 1 (b) and Figure 1 (d), we show the phase – voltage hysteresis obtained at $300K$ and $420K$ respectively at three different points. From a visual inspection of the data, it is clear that the phase switches by $180^{o}$ with a coercive voltage of approximately $10V$. The switching is sharp at $10V$ beyond which the hysteresis curve saturates.  As shown in Figure 1(c) and Figure 1(e), the amplitude vs. $V_{dc}$ curves also show hysteresis and form ``butterfly loop" at $300K$ and $420K$ respectively at three different points. This is believed to be the signature of piezoelectricity. By analysing the data, we have extracted the value of the piezoelectric coefficient which is $‘0.422’nm/V$ at $300K$. For temperatures above $420K$, no piezoelectric or ferroelectric like response was observed on SrTiO$_{3}$.

As it has been discussed before, based on the observation of hysteresis curves alone the ferroelectric phase of SrTiO$_{3}$ cannot be confirmed. Therefore we have attempted to write ferroelectric domains on SrTiO$_{3}$ using a conducting cantilever. A tip made of Silicon coated with $5nm$ of titanium and $20nm$ of iridium and spring constant $2N/m$ was used to write a circular domain using lithography. A circular domain was written by applying $+30V$ on $11\mu m \times 11\mu m$ area on the surface of bulk SrTiO$_{3}$ crystal. After writing the domain the area was scanned in regular PFM imaging mode. The height image did not show any modification while a bright circular domain was clearly visible in the phase image (Figure 2(a)). After that we applied a negative voltage ($-30V$) on a smaller concentric circular area inside the bright domain. When we imaged the domain again after the second writing, we observed a dark spot at the center of the bright domain. The phase difference between the bright region and the dark region was $180^o$ indicating that the polarization at the central dark region was reversed due to the application of a negative voltage. Therefore, we have successfully written and erased a ferroelectric domain on the surface of bulk SrTiO$_{3}$ crystal.

We also attempted to image the natural ferroelectric domains by regular and vector PFM. However, distinct domain structures were not observed. It is possible that the natural domains do exist but we cannot image them by PFM due to large tip size ($\sim$ 25nm) that limits the lateral resolution of the images.

In order to estimate the Curie temperature of the ferroelectric phase, temperature dependence of an artificially written circular domain was studied. We first wrote the circular domain of diameter 6 $\mu m$ by PFM lithography with an applied voltage of $+30V$ on SrTiO$_{3}$ at $300K$. The domain was then imaged using DART PFM at different temperatures as shown in Figure 3: $300K$ (Figure 3(a)), $330K$ (Figure 3(b)), $360K$ (Figure 3(c)), $390K$ (Figure 3(d)) and $420K$ (Figure 3(e)). From visual inspection alone it is clear that with increase in temperature the domain structure written at $300K$ starts disappearing, the diameter shrinks, and it disappears completely at $420K$ (also see supplementary material\cite{EPAPS}). After imaging the domain at every temperature we have also performed spectroscopic measurements in order to investigate the evolution of the relevant parameters like the coercive field and the piezoelectric constant. The spectroscopic data are presented as insets in Figure 3. Surprisingly, no systematic temperature dependence of the ferroelectric/piezoelectric parameters was noticed. However, no spectroscopic piezo and ferroelectric response on temperature above $440K$ could be found.

In order to investigate the time dynamics of the domains we have also measured the relaxation time (the time over which the domain completely disappears) of the domains. The domain survives for 12-14 hours at room temperature (see supplementary material).  As the temperature was increased from $300K$ up to $420K$, the relaxation time significantly decreases from 12-14 hours to 2-3 minutes. It was possible to write a domain even at $440 K$ but due to very short lifetime of the domain at this temperature the domain relaxed before it could be imaged. However, when we wrote the domain at $440 K$ and immediately cooled the sample down, the domain did not fully relax and it was possible to image the domain at room temperature (Figure 3(f)).

From the data presented above it is clear that the surface of SrTiO$_{3}$ is electrically active and the electrical response is hysteretic. In this context it should be noted that in the past the controlled bilayers of SrTiO$_{3}$ and   LaAlO$_{3}$ were also shown to be electrically active where the hysteresis effects were observed and electric domains were written.\cite{Xie,Bark,Huang} Such behaviour was attributed to the existence of an exotic interface in the SrTiO$_{3}$/LaAlO$_{3}$ bilayers. Since we observe similar effects on the surface of bulk SrTiO$_{3}$ alone, the origin of such effects in the SrTiO$_{3}$/LaAlO$_{3}$ bilayers should be revisited.

The surprising observation of very high temperature ferroelectricity in pure crystals of SrTiO$_{3}$ is not understood at present and further theoretical investigation is required. Qualitatively, the observation may be attributed to a voltage induced strain\cite{Pertsev} on the surface developed during the spectroscopic measurement and domain writing. When a DC voltage is applied on the surface through the conductive tip, the effective electric field is very large. At such high electric fields the lattice may distort in the direction of the applied field due to electrostriction. This distortion results in a change in the Ti-O bond length. This change breaks the inherent centrosymmetric nature of the cubic crystal structure of SrTiO$_{3}$ and induces electric polarization. It is not surprising that the lifetime of the induced strain is temperature dependent as at higher temperature the field-induced distortion should relax faster. The important thing to note, however, here is that the induced polarization has a rather long lifetime at room temperature. The microscopic mechanism leading to such long relaxation time should be investigated.

In conclusion, we have performed piezoresponse force microscopy on (110) single crystals of SrTiO$_{3}$ and found the evidence of a ferroelectric phase up to $440 K$. It was possible to write and erase electric domains using conducting cantilevers on the surface of the crystals. From the temperature dependence of the hysteresis effects and the electric domains we conclude that the ferroelectric Curie temperature is more than $440 K$. Based on the results presented here the origin of the exotic electric behaviour of the SrTiO$_{3}$/LaAlO$_{3}$ interfaces should be revisited.

We thank Dr. Jagmeet sekhon for his help. GS acknowledges the research grant of Ramanujan Fellowship from Department of science and Technology (DST), India for partial financial support. We also acknowledge fruitful discussions with Prof. Pushan Ayyub.


\begin{thebibliography}{99}

\bibitem{Dawber}M. Dawber, C. Lichtensteiger, M. Cantoni, M. Veithen, P. Ghosez, K. Johnston, K. M. Rabe, J.-M. Triscone, Phys. Rev. Lett. \textbf{95,} 177601 (2005).

\bibitem{Ohtomo} A. Ohtomo, H. Y. Hwang, Nature \textbf{427,} 423 (2004).

\bibitem{Ahn} C. H. Ahn, K. M. Rabe, J.-M. Triscone, Science \textbf{303,} 488 (2004).

\bibitem{Fong} D. D. Fong, G. B. Stephenson, S. K. Streiffer, J. A. Eastman, O. Auciello, P. H. Fuoss, C. Thompson, Science \textbf{304,} 1650 (2004).


\bibitem{Burnside} Shelly Burnside, Jacques-E. Moser, Keith Brooks, Michael Gratzel, D. Cahen, J. Phys. Chem. B \textbf{103,} 9328 (1999).


\bibitem{Tumarkin} A. V. Tumarkina, M. M. Gaidukova, A. G. Gagarina, T. B. Samoilovaa, A. B. Kozyreva, Ferroelectrics \textbf{439,} 49 (2012).

\bibitem{Manan}  M.M. Mehta,	D.A. Dikin,	C.W. Bark,	S. Ryu,	C.M. Folkman,	C.B. Eom	, V. Chandrasekhar, Nature Communications. \textbf{3}, 955 (2012).


\bibitem{Rimai} L. Rimai and G.deMars,Phys.Rev. \textbf{127}, 702 (1962).

\bibitem{Bell} R.O. Bell and G.Rupprecht, Phys.Rev.
\textbf{125},1915 (1962).

\bibitem{Zhong} W. Zhong, D. Vanderbilt, Phys. Rev. B. \textbf{53,} 5047 (1996).


\bibitem{Antons} A. Antons, J. B. Neaton, Karin M. Rabe, David Vanderbilt, Phys. Rev. B. \textbf{71,} 024102 (2005).

\bibitem{Muller} K. A. M\"{u}ller and H. Burkard, Phys. Rev. B \textbf{19}, 3593 (1979)

\bibitem{Itoh} M. Itoh, R. Wang, Y. Inaguma, T. Yamaguchi, Y-J. Shan, T. Nakamura, Phys. Rev. Lett. \textbf{82,} 3540 (1999).

\bibitem{Haenl} J. H. Haeni, P. Irvin, W. Chang, R. Uecker, P. Reiche, Y. L. Li, S. Choudhury, W. Tian, M. E. Hawley, B. Craigo, A. K. Tagantsev, X. Q. Pan, S. K. Streiffer, L. Q. Chen, S. W. Kirchoefer, J. Levy, D. G. Schlom, Nature \textbf{430,} 758 (2004).


\bibitem{Bilani-Zeneli} O. Bilani-Zeneli, A. D. Rata, A. Herklotz, O. Mieth, L. M. Eng, L. Schultz, M. D. Biegalski, H. M. Christen, K. Dorr, Appl. Phys. Lett. \textbf{104,} 054108 (2008).


\bibitem{Kim} Y. S. kim, D. J. Kim, T. H. Kim, T. W. Noh, J. S. Choi, B. H. Park, J. G. Yoon, Appl. Phys. Lett. \textbf{91,} 042908 (2007).

\bibitem{Maeng} W. J. Maeng, I. Jung, J. Y. Son, Solid state communications \textbf{152,} 1256 (2012).


\bibitem{Jang} H.W. Jang, A. Kumar, S. Denev, M.D. Biegalski, P. Maksymovych, C.W. Bark, C.T. Nelson, C.M. Folkman, S.H. Baek, N. Balke, C.M. Brooks, D.A. Tenne, D.G. Schlom, L.Q. Chen, X.Q. Pan, S.V. Kalinin, V. Gopalan, C.B. Eom, Phys. Rev. Lett. \textbf{104,} 197601 (2010).

\bibitem{Ravikumar}	V. Ravikumar, D. Wolf, V. P. Dravid, Phys. Rev. Lett. \textbf{74,} 960 (1995).

\bibitem{Bickel}	N. Bickel, G. Schmidt, K. Heinz, K. Muller, Phys. Rev. Lett. \textbf{62,} 2009 (1989).

\bibitem{Herger} R. Herger, P. R. Willmott, O. Bunk, C. M. Schleputz, B. D. Patterson, Phys. Rev. Lett. \textbf{98,} 076102 (2007).

\bibitem{Sekhon} J. S. Sekhon, L. Aggarwal, G. Sheet, e-print arXiv: 1401.2512v1.
\bibitem{Proksch} R. Proksch, e-print arXiv:1312.6933v1.


\bibitem{JesseA} S. Jesse, B. Mirman, S. V. Kalinin, Appl. Phys. Lett. \textbf{89,} 022906 (2006).

\bibitem{JesseB} S. Jesse, A. P. Baddorf, S. V. Kalinin, Appl. Phys. Lett. \textbf{88,} 062908 (2006).


\bibitem{EPAPS} Supplementary material.

\bibitem{Xie} Yanwu Xie, Christopher Bell, Takeaki Yajima, Yasuyuki Hikita, Harold Y. Hwang, Nano Lett. \textbf{10,} 2588 (2010).

\bibitem{Bark} C. W. Bark, P. Sharma, Y. Wang, S. H. Baek, S. Lee, S. Ryu, C. M. Folkman, T. R. Paudel, A. Kumar, S. V. Kalinin, A. Sokolov, E. Y. Tsymbal, M. S. Rzchowski, A. Gruverman, C. B. Eom, Nano Lett. \textbf{12,} 1765 (2012).


\bibitem{Huang} Mengchen Huang, Feng Bi, Chung-Wung Bark, Sangwoo Ryu, Kwang-Hwan Cho, Chang-Beom Eom, Jeremy Levy, e-print arXiv:1208.2687.

\bibitem{Pertsev} N. A. Pertsev, A. K. Tagantsev and N. Setter, Phys. Rev. B \textbf{61}, R825 (2000).



\end{thebibliography}
\end{document}